\documentclass[12pt]{iopart}
\usepackage{graphicx}
\usepackage{iopams}

\begin{document}

\bibliographystyle{iopart-num}

\title[Critiquing HF for the Anderson-Hubbard Model]
{Critiquing Variational Theories of the Anderson-Hubbard Model: 
Real-Space Self-Consistent Hartree-Fock Solutions}

\author{X. Chen$^1$, A. Farhoodfar$^1$, T. McIntosh$^1$, R. J. Gooding$^{1,3}$, P.W. Leung$^2$}
\address{$^1$Department of Physics, Engineering Physics and Astronomy, Queen's University, Kingston ON K7L 3N6
Canada}
\address{$^2$Department of Physics, Hong Kong University of Science and Technology, Clear Water Bay, Hong Kong}
\ead{$^3$gooding@physics.queensu.ca}
\begin{abstract}
A simple and commonly employed approximate technique with which one can examine spatially
disordered systems when strong electronic correlations are present is based on the use of 
real-space unrestricted self-consistent Hartree-Fock wave functions. In such an approach
the disorder is treated exactly while the correlations are treated approximately.
In this report we critique the success of
this approximation by making comparisons between such solutions and the exact wave
functions for the Anderson-Hubbard model. Due to the sizes of the complete Hilbert
spaces for these problems, the comparisons are restricted to small 
one-dimensional chains, up to ten sites, and a 4x4 two-dimensional cluster, and
at 1/2 filling these Hilbert spaces contain about 63,500 and 166 million states, 
respectively. We have completed these calculations both at and away from 1/2 filling.
This approximation is based on a variational approach which minimizes the
Hartree-Fock energy, and we have completed comparisons of the exact and
Hartree-Fock energies. However, in order to assess the success of this approximation 
in reproducing ground-state correlations we have completed comparisons of the local 
charge and spin correlations, including the calculation of the overlap of the 
Hartree-Fock wave functions with those of the exact solutions. We find that
this approximation reproduces the local charge densities to quite a high accuracy,
but that the local spin correlations, as represented by $\langle {\bf S}_i\cdot{\bf S}_j\rangle$,
are not as well represented. In addition to these comparisons, 
we discuss the properties of the spin degrees of freedom in the HF approximation,
and where in the disorder-interaction phase diagram such physics may be important.
\end{abstract}

\pacs{71.27.+a,71.23.-k,71.10.-w}

\maketitle

\section{Introduction}\label{sec:intro}

Many transition metal oxides display phenomena that are believed to be associated with
both strong electronic correlations and disorder. Various metal-to-nonmetal
transitions \cite{imada98}, and some properties of the underdoped high-T$_c$ cuprate 
superconductors \cite{kastner98,gooding97,lai98} are examples of such physics. 
Presently the study of such systems is a very active field of research in condensed matter
physics.

Theoretically, the simplest model that hopefully represents some of the key physics
of such systems is the so-called Anderson-Hubbard Hamiltonian. The Anderson 
model \cite{anderson58} is given by
\begin{equation}
\label{eq:Ham_AndHub}
{\hat H_A}~=~\sum_{i,\sigma}~V_i~{\hat n}_{i,\sigma}~-~t~\sum_{\langle i,j \rangle,
\sigma}~\Big({\hat c}^\dagger_{i,\sigma}~{\hat c}_{j,\sigma}~+~h.c.\Big)
\end{equation}
where $i,j=1\dots N$ denote the sites of the lattice, $\langle i,j \rangle$ implies
that $i$ and $j$ are near neighbours, ${\hat c}_{i,\sigma}$ (${\hat n}_{i,\sigma}$) 
is the destruction (number) operator for an electron at site $i$ with spin $\sigma$,
and the hopping energy is denoted by $t$. The on-site energy at site $i$ is given by 
$V_i$, and in this report we have examined two models of disorder: (i) a 50/50 binary 
alloy model, where $V_i$ is set equal to $W/2$ (for an $A$ site) or $-W/2$ (for a $B$ 
site); and (ii) a box distribution of on-site energies in which all site
energies in the range $-W/2$ to $+W/2$ are equally likely. The electron interactions 
that are included are represented by the Hubbard term \cite{hubbard63}, given by
\begin{equation}
\label{eq:Ham_Hub}
{\hat H_H}~=~U~\sum_i~{\hat n}_{i,\uparrow}{\hat n}_{i,\downarrow}
\end{equation}
The Anderson-Hubbard model is formed from the sum of ${\hat H_A}$ and ${\hat H_H}$.

There has been considerable work on this model over the last several decades, and
most of the interest in this model arises from the hope that this model contains some
of the essential physics required to describe disordered transition-metal 
oxides, including their metal-to-insulator transitions \cite{imada98}. 
For example, several numerical calculations have been completed using different variants
of the quantum Monte-Carlo technique \cite{ulmke97,dentenner99,otsuka00,enralran01,paris07}.
There are also some exact solutions available for lattices of very small 
sizes \cite{benenti99,kotlyar01,berkovits01,berkovits03,srinivasan03,vasseur05},
including very recent work on the exact spectral functions for lattices of 10 sites
\cite{chiesa08}. 

Here we present our exact results for several one-dimensional chains and
a two-dimensional 4x4 lattice at and below half filling. In this, and a future 
companion paper \cite{farhoodfar08} introducing the application of partially-projected
Gutzwiller variational 
wave functions to the Anderson-Hubbard model, we use these solutions to assess the
ability of variational wave functions to represent both the energies and
correlations of ground-state wave functions. Here we present our comparisons
of the exact solutions to those found in the Hartree-Fock (HF) approximation.

Indeed, the simplest and most common approach to finding approximate solutions for interacting
systems is that of using a self-consistent, real-space unrestricted Hartree-Fock approach.
In this approach the disorder is treated exactly, thus allowing for the inclusion of coherent
back scattering, whereas the interactions are dealt with in an approximate fashion.
The HF approach has been used in various ways in previous studies, and some examples
include (i) an attempt to study the high-$T_c$ cuprates via a determination of the phase
diagram of the disordered two-dimensional Hubbard model \cite{dasgupta93}, (ii) a detailed
examination of the 3d metal-to-insulator transition \cite{tusch93}, including
a discussion of transitions in some cubic disordered tungsten bronzes \cite{duecker99}, 
(iii) a proposal for a novel metallic phase in two dimensions that results from the combined effects
of correlations and disorder \cite{heidarian04,trivedi05a,trivedi05b}, and (iv) work by one of us and 
coworkers \cite{fazileh06} studying the metal-to-insulator transition in  LiAl$_y$Ti$_{2-y}$O$_4$,
examining the possibility of the presence of strong correlations in the LiTi$_2$O$_4$ system
(which has been proposed \cite{mueller96} to be a non-cuprate fully three-dimensional material that is
related to the quasi-two-dimensional high-$T_c$ cuprates). The conclusions drawn from
these studies are subject to the veracity of the approximation used, namely the HF approach,
and a focus of this paper is to provide a partial assessment of this technique. 

\section{Real-Space Self-Consistent Hartree-Fock Approximation}

The HF decoupling scheme is well known, discussed at length elsewhere (\textit{e.g.}, see
\cite{auerbachbk}), but for completeness we summarize the final results. The HF technique
can be thought of as an approach in which one ignores terms in the interaction Hamiltonian,
\textit{viz.} the Hubbard Hamiltonian of Eq.~(\ref{eq:Ham_Hub}), that are proportional 
to fluctuations about mean values squared. That is, one approximates the local Hubbard 
interaction as being replaced by (here we ignore the possibility of local superconducting 
pairing correlations)
\begin{eqnarray}
\label{eq:2siteH}
\langle {\hat n}_{i,\uparrow}{\hat n}_{i,\downarrow} \rangle 
&=& ({\overline n}_{i,\uparrow}
+\delta {\hat n}_{i,\uparrow})({\overline n}_{i,\downarrow} +\delta {\hat n}_{i,\downarrow}) 
- (h^+_i+\delta {\hat h^+}_i) 
(h^-_i+\delta {\hat h^-}_i)\\ \nonumber
&\approx&{\hat n}_{i,\uparrow} {\overline n}_{i,\downarrow} +
{\hat n}_{i,\downarrow} {\overline n}_{i,\uparrow} -
{\overline n}_{i,\uparrow} {\overline n}_{i,\downarrow} -
{\hat h^+}_i h^-_i -
{\hat h^-}_i h^+_i +
h^+_i h^-_i 
\end{eqnarray}
where ${\overline n}_{i,\sigma}\equiv\langle{\hat n}_{i,\sigma}\rangle$, and the effective local 
fields $h_i^\pm$ are given by \begin{equation}
h_i^+ \equiv \langle {\hat S^+}_i \rangle~~~~~~\textrm{and}~~~~~~~~~~
h_i^- \equiv \langle {\hat S^-}_i \rangle~~.
\end{equation}
Then, one must find self consistently the local spin-resolved charge densities and local fields
that minimize the variational estimate of the ground-state energy.

One aspect of our study is to examine the manner in which allowing for paramagnetic (PM) HF solutions,
\textit{i.e.} ${\overline n}_{i,\uparrow}~=~{\overline n}_{i,\downarrow}$ and $h_i^+=h_i^-=0$,
in comparison to restricted HF solutions, \textit{i.e.} 
${\overline n}_{i,\uparrow}~\not=~{\overline n}_{i,\downarrow}$ (in general) and $h_i^+=h_i^-=0$,
which from now on we refer to as the AFM HF solutions,
in comparison to fully unrestricted HF solutions,\textit{i.e.}
${\overline n}_{i,\uparrow}~\not=~{\overline n}_{i,\downarrow}$ (in general) and $h_i^+\not=h_i^-\not=0$
(in general) improves the correct representation of the correlations in these systems. That is,
by allowing for new degrees of freedom the variational principle guarantees that we can only lower
the variational estimate of the ground-state energy. Indeed, we present results of such
behaviour in this paper. However, while one has improved on the estimate of the energy, does that
necessarily mean that one is doing a better job at representing the correlations using such
variational wave functions? After presenting our results we will discuss them in relation to 
certain results in the literature \cite{tusch93,fazileh06} that rely on choosing which magnetic
degrees of freedom are ``active" in their associated variational wave functions. 

\begin{figure}[htbp]
\begin{center}
\includegraphics[height=8.6cm,width=6.75cm]{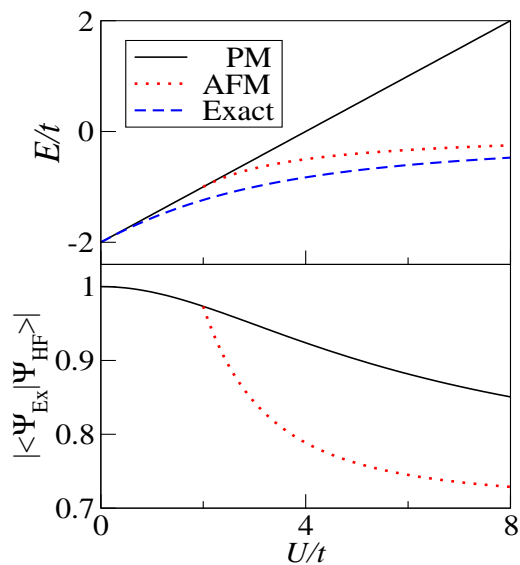}
\caption{\label{fig:2siteordered}
The upper panel shows a comparison of the exact ground-state energy of a two-site
ordered cluster with paramagnetic (PM) and non-paramagnetic antiferromagnetic (AFM)
solutions as a function of $U/t$. The lower panel shows a similar comparison for the overlap 
quantity defined by $|\langle \Psi_{\textrm{ex}} | \Psi_{\textrm{HF}} \rangle|$.  
Although the AFM energy is closer to the exact energy for $U/t > 2$ the overlap with the PM HF solution
is better over the same range of interactions.
}
\end{center}
\end{figure}

As a demonstration of this issue in figure \ref{fig:2siteordered} we show both the AFM and PM
HF energies, and the magnitudes of the overlaps of these wave functions with those of the exact
solutions, for a two-site (ordered) cluster (with open boundary conditions). One sees that for $U/t>2$ the
AFM solutions have lower energies, but (unfortunately) the overlaps of the PM and exact wave functions
are far larger than the overlaps of the AFM and exact wave functions in this same range of $U/t$.
That is, obtaining a better variational estimate of the energy does not guarantee a better representation
of the ground-state wave function.

The development of the magnetic ``order", which leads to the lowering of the HF energy for
$U/t > 2$, is associated with $\frac{1}{2}({\overline n}_{i,\uparrow}~-~{\overline n}_{i,\downarrow})$
becoming nonzero. However, like the overlap shown above, this does not lead to an improved representation of
the spin correlations present. For the exact ground state, as $U/t\rightarrow\infty$ one finds
$\langle \Psi | {\hat {\bf S}}_1\cdot {\hat {\bf S}}_2  | \Psi\rangle\rightarrow-\frac{3}{4}$. The PM
HF solution gives $-\frac{3}{8}$ for all $U/t>0$, whereas for the AFM HF solution at large $U/t$ one
finds essentially anti-parallel classical spins, thus giving a value of $-\frac{1}{4}$.
Therefore, at least in this one example, as one improves the energy by allowing for magnetic HF 
solutions one in fact decreases the agreement between the exact and HF spin correlations.

For spatially disordered systems all calculations must be computed numerically. We have found all
self-consistent HF ground-state wave functions to an absolute accuracy (for the local spin-resolved
charge densities and effective fields) of at least 10$^{-5}$, and often to a much higher accuracy.

\section{Comparisons of Exact and HF Quantities}

The quantities that we have calculated, in addition to the exact and HF variational ground-state energies, 
are as follows. 
\begin{itemize}
\item{}We have evaluated the magnitude of the overlap between
the exact and HF variational wave functions, given by
\begin{equation}
|\langle \Psi_{\textrm{ex}} | \Psi_{\textrm{HF}} \rangle|~~.
\end{equation}
\item{} We have calculated the local charge densities according to
\begin{equation}
\label{eq:localCDdefn}
{\overline n}_i \equiv \langle \Psi | \sum_\sigma {\hat n}_{i,\sigma} | \Psi \rangle~~.
\end{equation}
\item{} We have calculated the local spin exchanges for near-neighbour sites according to
\begin{equation}
\label{eq:SdotSdefn}
\langle \Psi | {\hat {\bf S}}_i\cdot
{\hat {\bf S}}_j  | \Psi \rangle~~.
\end{equation}
\end{itemize}
For a small cluster of only 4 sites it is in fact very helpful to show the charge densities
for all sites. However, for larger systems this is not as practical to do this,
and instead we have calculated the average absolute difference of the exact and HF charge densities,
defined by
\begin{equation}
\label{eq:msqdiff-rhos}
\langle \delta {\overline n}\rangle=\frac{1}{N}\sum_{i=1}^N~|{\overline n}_i^{~var}-{\overline n}_i^{~exact}|~~.
\end{equation}

\subsection{Results: One-Dimensional Chains}\label{sec:resultsin1d}

\subsubsection{Four Site Cluster}

\begin{figure}[htbp]
\begin{center}
\includegraphics[height=7cm,width=6.75cm]{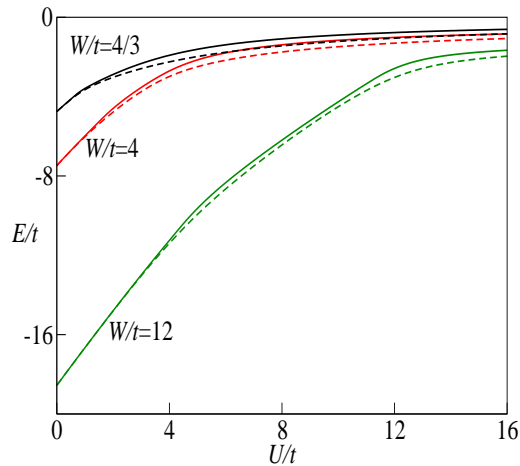}
\caption{\label{fig:4sitenrgs}
A comparison of the exact and HF energies for the four-site disordered ($W/t = 4/3,~4,~12$) cluster
with the disorder as given by Eq.~(\protect\ref{eq:4sitedisorder}). 
For all colours, the (lower) dashed curve is the exact energy and the solid line is
the RSSCHF energy. As expected, for large correlations ($U \gg t,W$) all energies approach
the same value.
}
\end{center}
\end{figure}

\begin{figure}[htbp]
\begin{center}
\includegraphics[height=8cm,width=7.25cm]{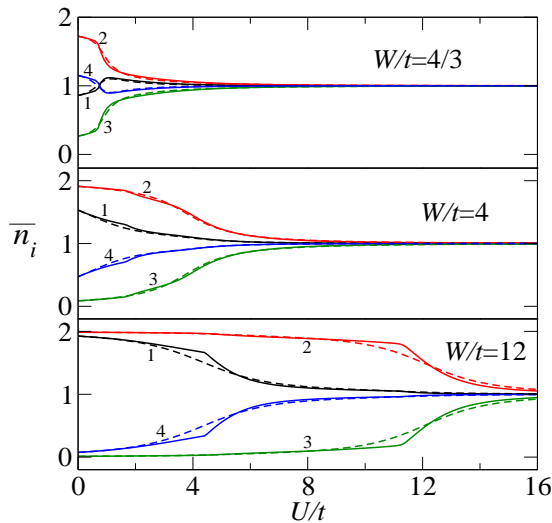}
\caption{\label{fig:4sitenis}
A comparison of the exact and HF local charge densities for the four-site disordered
($W/t = 4/3,~4,~12$) cluster with the disorder as given by Eq.~(\protect\ref{eq:4sitedisorder}).
The numbers 1, 2, 3 and 4 label the different lattice sites.
For all colours, the dashed curve is the exact result and the solid line is
the HF result. Clearly, the agreement of the exact and HF results is very good.
}
\end{center}
\end{figure}

\begin{figure}[htbp]
\begin{center}
\includegraphics[height=8cm,width=6.75cm]{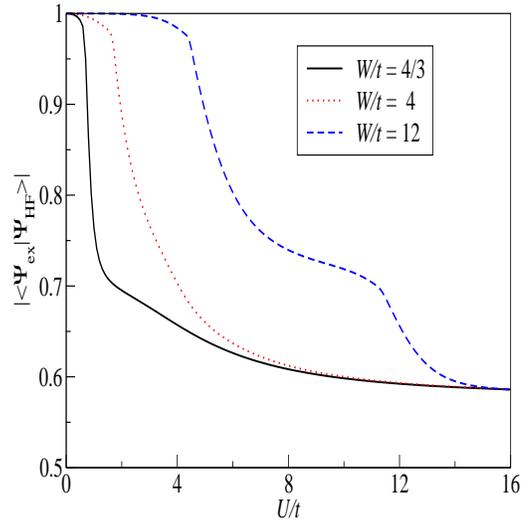}
\caption{\label{fig:4siteoverlap}
The magnitude of the overlap of the exact and HF wave functions
for the four-site disordered ($W/t = 4/3,~4,~12$) cluster.
}
\end{center}
\end{figure}

\begin{figure}[htbp]
\begin{center}
\includegraphics[height=10.5cm,width=8cm]{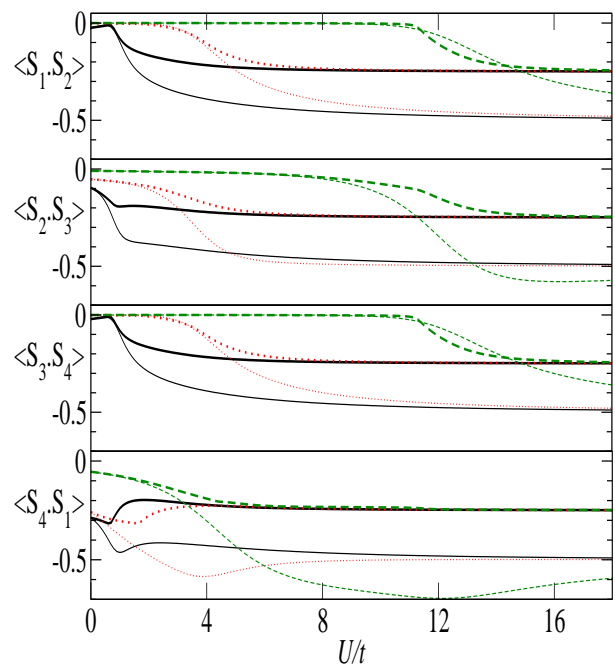}
\caption{\label{fig:4siteSidotSj}
A comparison of the exact and HF expectation values of
$\langle \Psi | {\hat {\bf S}}_i\cdot {\hat {\bf S}}_j  | \Psi \rangle$ for the four-site disordered 
cluster with the disorder as given by Eq.~(\protect\ref{eq:4sitedisorder})
($W/t = 4/3$ (solid lines), 4 (dotted lines), 12 (dashed lines)). For all colours, the thinner curve 
is the exact result and the thicker line is the HF result. 
}
\end{center}
\end{figure}

We have examined several complexions of disorder for the $4\times 1$ cluster at 1/2 filling, 
but as mentioned above for this small cluster it is instructive to focus on one representative 
configuration. The disorder for the configuration discussed below is given by
\begin{equation}
\label{eq:4sitedisorder}
V_i~=~-W/4,~-W/2,~+W/2,~+W/6~~~~~~~\textrm{for}~~~~ i = 1~~\textrm{to}~~4.
\end{equation}
The ``bandwidth" for the noninteracting ordered $4\times 1$ cluster
(with periodic boundary conditions) is $4t$, and therefore we have examined weak $W/t = 4/3$,
intermediate $W/t = 4$ and strong $W/t = 12$ disorder, and then varied $U/t$ continuously
from zero to roughly twice $W/t$. Our results are shown in figures
\ref{fig:4sitenrgs} to \ref{fig:4siteSidotSj}.

Referring to figure \ref{fig:4sitenrgs}, one sees that for all disorder 
and interaction strengths the energies are quite similar. If one
use non-magnetic PM HF solutions one obtains energies that are not at all similar to the
exact energies -- in fact, the PM energies increase linearly with $U/t$, similar to the 
two-site results shown in figure \ref{fig:2siteordered}.

As seen in figure \ref{fig:4sitenis} the local charge densities are very close to the exact values, 
and as we show below this is true for all clusters that we have studied. This supports the 
rubric that HF works very well for 
disordered systems, but more specifically HF is impressive in its abilities to reproduce
the inhomogeneous charge densities found in interacting disordered electronic systems. 
The charge densities obtained from PM HF solutions
look nothing like the exact solutions, and, in fact, do not become uniform in the limit
of large $U/t$. Similarly, very poor comparisons of the PM HF results with other
correlation functions are found, and for the rest of this report we limit our attention
to (potentially) magnetic HF ground states.

The unrestricted HF solutions for this cluster lead to magnetic transitions to states with collinear
spins, meaning that the effective fields $h^\pm_i$ are equal to zero. 
For weak and intermediate disorder these transitions, which occur at $U/t \approx 0.6$ and 1.6, respectively,
lead to large moments (defined for collinear spins as 
$\frac{1}{2}(\overline{n}_{i\uparrow} - \overline{n}_{i\uparrow})$), while for strong disorder two transitions 
to states with collinear spins are found -- first sites 1 and 4 develop large moments while sites
2 and 3 develop moments that are small and then decrease to zero, and for larger $U/t$ sites 2 
and 3 also develop larger moments. (We show plots of similar data for the ten-site cluster in 
the next section.) Of course, these transitions are spurious, as the exact ground states are singlets 
for all disorders and $U/t>0$ and no moments are present on any site. 
Nonetheless, allowing for magnetic HF solutions is important in finding
ground-state energies and local charge densities that are in good agreement with the exact values.

The overlaps with the exact wave functions shown in figure \ref{fig:4siteoverlap} are very close to one 
for $U/t$ less than 0.6, 1.6 and 4.3, for weak, intermediate and strong disorder (the numerical 
values of the overlaps are in excess of 0.975). However, for larger values of $U/t$ the HF wave
functions are reduced to 60-70 \%. Noting that these are the values of disorder and Hubbard
energy at which magnetic moments first develop, we see that the very good overlap with
the exact solutions is obtained only for PM HF solutions -- for comparison, 
see figure \ref{fig:2siteordered}. By searching through the numerical values of the probability
amplitudes of the exact solutions we have determined that these substantial overlaps 
(for $U/t$ less than 0.6, 1.6 and 4.3, for weak, intermediate and strong disorder) are largely 
due to the fact that the exact ground state is dominated by one single probability amplitude, 
and therefore a product-state solution such as HF is able to reliably reproduce such a state. We 
find that this trend persists even when the Hilbert spaces are much larger (for larger clusters 
at various electronic fillings).

In figure \ref{fig:4siteSidotSj} we show $\langle{\hat {\bf S}}_i\cdot{\hat {\bf S}}_j\rangle$ for
each near-neighbour pair of sites, and similar to the overlaps we find reasonably good
agreement for small $U/t$ (relative to $W/t$), but poor agreement at larger Hubbard energies.
The failure of the product-state-based HF solution to represent the fluctuating quantum
spins found in the large $U/t$ is the same as discussed earlier for the ordered two-site problem.

\subsubsection{Other Chains}

The data for the four-site cluster with one particular complexion of disorder are, in fact, very
similar to what we find for larger clusters. However, it is desirable to study larger systems with
much larger Hilbert spaces. That being the case, it would be good to be able to study larger
systems that do not suffer from finite-size effects that are too severe. To address this matter
we have studied $L=$ 4, 6, 8, and 10 chains with exact diagonalization, and have found that
quantities such as the charge density converge with increasing $L$ when sizes of 6 and 10
are viewed. This is also the case for the HF solutions, and for which we also find that the $L=14$
HF result is very similar to the $L=6$ and 10 HF results. Therefore, below we discuss our results
for $L=10$ chains with periodic boundary conditions. (This kind of finite-size behaviour has been 
noted in studies of ordered one-dimensional chains -- \textit{e.g.}, see \cite{kaplan82}.)

For larger systems it is not easy to view such large collections of data. Further, the previous 
subsection focused on only one complexion of disorder,
and it is preferable to examine such data that is subject to a configurational average over
various complexions of disorder. Here we show data at 1/2 filling and close to 1/4 filling
for a ten-site chain with periodic boundary conditions (10 and 6 electrons, respectively) that
are averaged over 10 different complexions of randomly chosen on-site energies, again for
weak, intermediate and strong disorder. Further, for the four-site cluster at half filling 
there are (4-choose-2-squared) 36 states in the complete Hilbert space, whereas for the 
ten-site clusters mentioned above the dimensionalities of the Hilbert spaces are 14,400 and 63,504 
for 6 and 10 electrons, respectively. We have found the exact solutions for these larger systems 
by using the Lanczos method.

\begin{figure}[htbp]
\begin{center}
\includegraphics[height=7.cm,width=7.5cm]{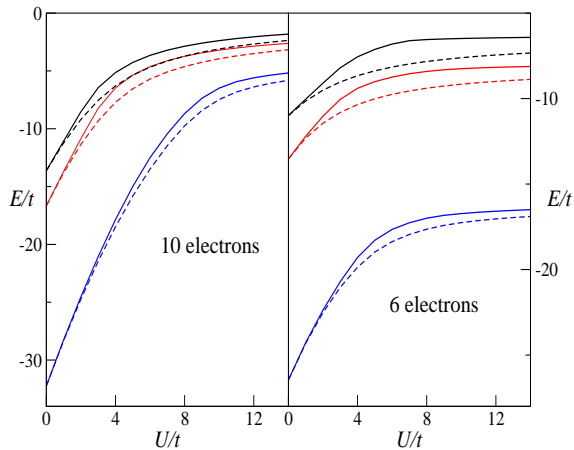}
\caption{\label{fig:nrgsL10Ne10and6}
A comparison of the exact and HF expectation values of the ground-state energies
the ten-site disordered ($W/t = 4/3,~4,~12$) cluster averaged over ten complexions
of disorder, for 1/2 filling (left panel) and close to 1/4 filling (right panel).
For all colours, the solid curve is the HF energy and the dashed line is
the energy of the exact ground state. 
}
\end{center}
\end{figure}

In figure \ref {fig:nrgsL10Ne10and6} we show the disorder-averaged ground-state energies of 
the exact and HF solutions.  While the agreement is not as good as found for four sites, the 
HF approximation does quite well in reproducing the energy.

\begin{figure}[htbp]
\begin{center}
\includegraphics[height=7.cm,width=9.5cm]{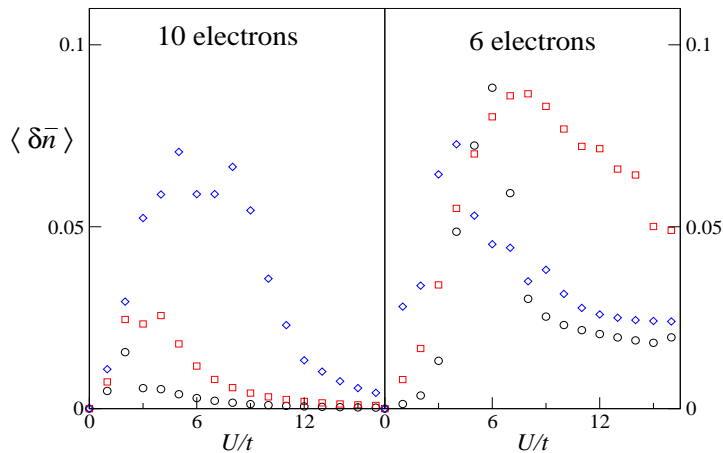}
\caption{\label{fig:niL10Ne10and6}
A comparison of the mean absolute difference between the exact and HF expectation values of the
local charge densities the ten-site disordered cluster averaged over
ten complexions of disorder, as described in the text, at 1/2 filling (left panel) and
close to 1/4 filling (right panel). The results are for disorder strengths of $W/t=4/3$ 
(black circles), $W/t=4$ (red squares) and $W/t=12$ (blue diamonds).
}
\end{center}
\end{figure}

In order to discuss the success of the HF approximation in reproducing the local charge
densities below we show the mean absolute difference of this quantity, defined  in
equation (\ref{eq:msqdiff-rhos}), nothing that we include an average of this quantity over
ten complexions of disorder. This quantity is shown in figure \ref{fig:niL10Ne10and6},
from which we see that at half filling and for weak and intermediate
disorder the HF approximation does very well at reproducing the local charge densities -- the largest
(disorder averaged) absolute difference is at most 2\%. However,
for strong disorder, until the exact ground states are found to become homogeneous at large $U/t$
the agreement with HF is not as impressive -- the largest
(disorder averaged) absolute difference can be of the order of 6-7\%. 
Away from 1/2 filling the agreement is found to be worse for all disorder strengths. However,
for all disorder and interaction strengths the disorder-averaged absolute difference is always 
less than 9\%.

\begin{figure}[htbp]
\begin{center}
\includegraphics[height=9.cm,width=11.cm]{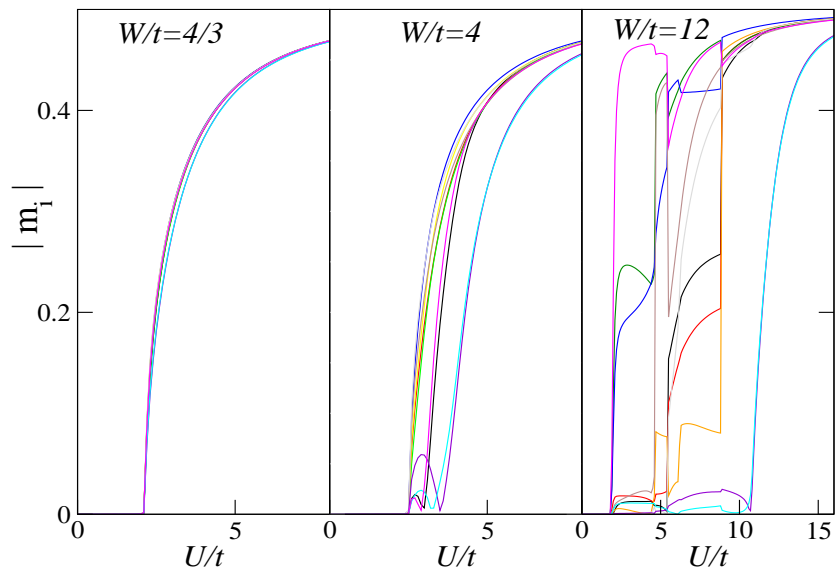}
\caption{\label{fig:momentsinL10Ne10and6}
The magnitudes of the magnetic moments, defined for collinear spins as 
$|$m$_i|=\frac{1}{2}(\overline{n}_{i\uparrow}
- \overline{n}_{i\downarrow})$, that develop in one particular complexion of disorder for 10 sites
and 10 electrons for weak (left panel), intermediate (centre panel) and strong (right
panel) disorder as a function of $U/t$. Each line type corresponds to one of the sites
of the lattice.
}
\end{center}
\end{figure}

The magnetic properties of these HF solutions are nontrivial. At 1/2 filling and weak 
disorder all sites develop magnetic moments at the same value $U/t$ for a given complexion 
of disorder, and the disorder-averaged value is roughly $U/t\sim2.05$. However, for 
intermediate and strong disorder the situation is much more complicated, and using 
a representative complexion of disorder in figure \ref{fig:momentsinL10Ne10and6} we show
the local magnetic moments for increasing $U/t$ for all three strengths of disorder. 
Note the results shown in the middle panel are qualitatively similar to those discussed 
(but not shown) previously for the four-site cluster with strong disorder. 
Further, for certain (but not all) complexions of disorder for intermediate values
of $U/t$ (in the range of 4-10) we find ground states having \textit{coplanar non-collinear spins}. 
As we discuss below, there is also a small range of $U/t$ that we find such non-collinear 
spin textures in our two-dimensional HF results, and we defer the discussion of the physical 
significance of this result until we present that data.

If one studies the behaviour of $\langle \Psi | {\hat {\bf S}}_i\cdot {\hat {\bf S}}_j  | \Psi \rangle$
for the exact and HF results for 10-site chains for each complexion one finds results quite similar to 
those shown in figure \ref{fig:4siteSidotSj}. That is, for $U/t \lesssim W/t$ one finds reasonably close
agreement for some pairs of near-neighbour sites, but for larger $U/t$ the magnitudes of these 
quantities can be very different, with the HF result confined to always give -1/4 for large $U/t$.

\subsection{Results: Two-Dimensional Square Lattice}\label{sec:resultsin2d}

We now discuss our results for a two-dimensional square lattice. In contrast to
the earlier work, here we focus on an AB binary model of disorder in which $V_i=W/2$
for A sites and $V_i=-W/2$ for B sites (the spatial arrangement may
be identified by referring to figure \protect\ref{fig:spintwists} or table
\ref{table:SdotSVeq7Ueq7halffilling}). 
This work may be relevant to recent experiments on binary alloy monolayers \cite{pratzner03}.

In our numerical work we have studied a binary alloy for a 4$\times4$ lattice with 
periodic boundary conditions, with disorder strengths of $W/t=$ 4, 6, 7, 8, 9, 10, and 12,
and on-site Hubbard repulsion energies of $U/t=$0, 4, 8 and 12. We considered
8 (1/4 filling), 12 (3/8 filling), and 16 (1/2 filling) electrons. These energies
should be compared to the bandwidth of the ordered, non-interacting problem, which for this
lattice is 8$t$. Due to the lack of translational periodicity, one must determine
all of the 16-choose-8 squared (for 1/2 filling), or roughly 166 million probability amplitudes,
and to date there are no exact wave functions available for such a large spatially disordered system.
Due to the enormity of the Hilbert space we have considered only one complexion of (binary)
disorder, but as mentioned above have considered many disorder and interaction strengths.

This is a very large 
data set containing a considerable amount of interesting physics. However, since 
this paper is restricted to critiquing the HF approximation here we focus on a small 
subset of this data. Other data will be published along with a discussion of some
of the more interesting physics results that are found in this work at a later time.

\begin{figure}[htbp]
\begin{center}
\includegraphics[height=7.cm,width=7.cm]{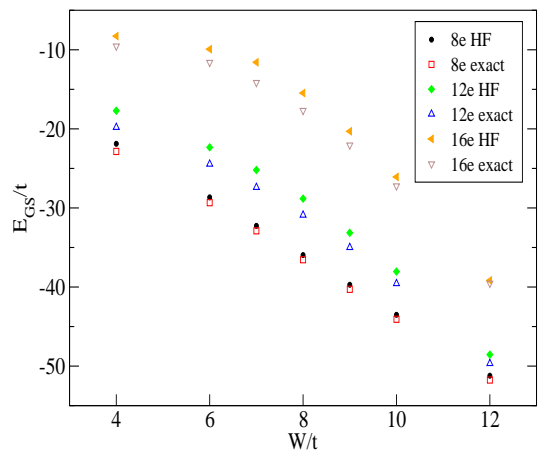}
\caption{\label{fig2:exactvsHFnrgs}
The exact {\em vs.} HF energies for the
binary alloy Hamiltonian discussed in the text. The Hubbard interaction
is fixed to be $U/t=8$, and the lower/middle/upper curves correspond to
fixed electronic densities of one-quarter/three-eighths/half filling, namely
8, 12 and 16 electrons (8e, 12e and 16e).
}
\end{center}
\end{figure}

\begin{figure}[htbp]
\begin{center}
\includegraphics[height=6.5cm,width=6.5cm,clip=true]{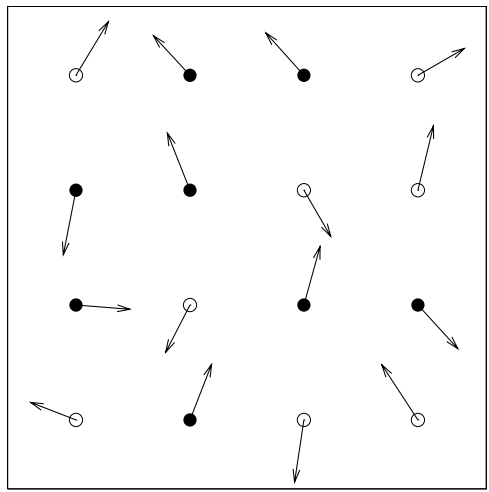}
\caption{\label{fig:spintwists}
The spin configuration for a 1/2 filled system with $W/t=7$ and $U/t=8$. The filled circles denote
A sites ($V_i=W/2$) and the open circles denote B sites ($V_i=-W/2$). The length of each vector is 
proportional to the magnitude of the magnetic moment on that site.
}
\end{center}
\end{figure}

A comparison of the energies for different fillings and disorder for $U/t=8$ is shown in 
figure \ref{fig2:exactvsHFnrgs} -- similar results are found for $U/t=$ 4 and 12. 
The agreement between the exact and HF energies is seen 
to improve with increasing disorder, which is what one expects. 
One finds parallel results if one calculates the overlap of the exact and HF wave
functions -- for all fillings the overlap is maximized for the largest disorder,
indicating that for this parameter regime the ground states are better approximated
by product states, {\em viz.}, the ground-state wave functions are close to those
found for an equivalent Anderson ground state. 

A very good agreement between the exact and HF local charge densities is found
for all parameter sets for the 4$\times$4 lattice, similar to what
was shown in the previous subsections in one dimension. To emphasize the success
of the HF approximation we consider the parameter region of strongest competing 
interactions, namely when all of the kinetic, disorder and interaction energies
are close to one another, since it is in this region that it should be most 
difficult for HF to provide accurate results. In table \ref{table:chargesVeq7halffilling} 
we show the local charge densities for $W/t=7$ and $U/t=8$ at 1/2 filling. One sees that 
this leads to an inhomogeneous state with an average charge density of roughly 0.6 and 1.4 on
the A and B sites, respectively, and, as seen from a comparison with the $W/t=7$ and $U=$0 
result, such charge densities are strongly influenced by the interactions. However, the 
interactions are not strong enough to produce a uniform charge density, such as
that seen in figure \ref{fig:4sitenis} for four sites at large $U/t$. The agreement of
the exact and HF states is very good: for one site the absolute difference is almost 8\%, 
whereas for all other sites the absolute difference is usually 3-4\% or less.

\begin{table}
\begin{tabular}{cccccccc}
&&(a)      \\[4pt]
&&&&1.918 & $\,$ 0.045 & $\,$ 0.113 & $\,$ 1.953 \\[2pt]
&&&&0.085 & $\,$ 0.071 & $\,$ 1.902 & $\,$ 1.915 \\[2pt]
&&&&0.070 & $\,$ 1.868 & $\,$ 0.086 & $\,$ 0.085 \\[2pt]
&&&&1.933 & $\,$ 0.100 & $\,$ 1.901 & $\,$ 1.954 \\[10pt]
&&(b)      \\[4pt]
&&&&1.333 & $\,$ 0.586 & $\,$ 0.580 & $\,$ 1.440 \\[2pt]
&&&&0.652 & $\,$ 0.654 & $\,$ 1.453 & $\,$ 1.327 \\[2pt]
&&&&0.550 & $\,$ 1.438 & $\,$ 0.641 & $\,$ 0.570 \\[2pt]
&&&&1.431 & $\,$ 0.626 & $\,$ 1.394 & $\,$ 1.328 \\[10pt]
&&(c)      \\[4pt]
&&&&1.372 & $\,$ 0.540 & $\,$ 0.581 & $\,$ 1.447 \\[2pt]
&&&&0.656 & $\,$ 0.623 & $\,$ 1.431 & $\,$ 1.347 \\[2pt]
&&&&0.567 & $\,$ 1.438 & $\,$ 0.618 & $\,$ 0.596 \\[2pt]
&&\qquad&\qquad&1.471 & $\,$ 0.607 & $\,$ 1.361 & $\,$ 1.346 \\
\\
\end{tabular}
\caption{\label{table:chargesVeq7halffilling}
In (a) the exact local charge densities are listed for
the $U/t=0$ non-interacting ground state at 1/2 filling for $W/t=7$; (b) shows analogous
data, but now for the exact ground state of the interacting system with $U/t=8$; (c) shows 
the local charge densities for the interacting problem, now calculated within the HF approximation.}
\end{table}

One obtains even better agreement in other parameter regions, and sometimes finds nearly perfect 
agreement. For example, at 1/4 filling (note that at this filling there are 8 low-energy B sites 
and 8 electrons) for $U/t=8$ and $W/t=12$, the largest absolute difference in charge densities
is 0.002. Clearly, HF is very successful in representing the correct local charge densities
for the ground state.

We now turn to what one can learn from the HF solutions concerning the spin degrees of
freedom for such systems, both the spin-spin correlations and the spatial arrangement
and orientations of the spins. The parameter set used in table \ref{table:chargesVeq7halffilling}
leads to a remarkable result. Out of all of the exact diagonalization results, all
ground states correspond to singlets except for $W/t=7$ and $U/t=8$, for which one finds
a triplet ground state. Further, for the HF ground states only for this parameter set does one
find a solution with a nonzero net moment, and this moment is equal to one! Further,
for this parameter set one finds \textit{non-collinear spins} -- for $h^\pm_i=0$ one finds
a HF energy of -11.52$t$ whereas for $h^\pm_i$ non-zero (and real) one finds a HF
energy of -11.58$t$ (the exact energy is -14.14$t$). Further, the agreement between
the HF and exact and local charge densities is better for the non-zero $h^\pm_i$ HF
state than the HF state with collinear spins, so not only does one get a lower energy
but one also obtains better charge correlations! Since the spins in this HF state are 
non-collinear the HF ground state corresponds to a ``twisted spin configuration", and in 
figure \ref{fig:spintwists} we show the spin texture that we find. Clearly, this is not
a simple antiferromagnetic ground state with anti-parallel spins.

As discussed previously for one dimension, the effectiveness of the HF approximation 
in reproducing the spin correlations in the ground state is not expected to be as good
as for the local charge densities, and indeed this is what we find. There is only a 
reasonable {\em qualitative} similarity between the exact and HF near-neighbour spin-spin 
correlations. In the interests of completeness, \textit{e.g.} for comparisons made to
other approximation schemes, in table \ref{table:SdotSVeq7Ueq7halffilling} 
we show one typical example of this comparison,
and, in particular, for the same parameters as the above-displayed charge densities, 
namely 1/2 filling and $W/t=7$ and $U/t=8$. From this one sees that while many of 
the correlations found in the exact and HF solutions are similar to one another, 
and the largest value for the exact solution (-0.182) is not that dissimilar to the
HF solution\break (-0.125), sometimes HF even manages to get the wrong sign (\textit{e.g.},
-0.19 \textit{vs.} +0.045).
Further, comparing the exact and HF results using a spatially averaged absolute difference of
the near-neighbour spin-spin correlations, while the local charge densities improve with
increasing disorder, similar to the energies of figure \ref{fig2:exactvsHFnrgs} and the overlaps
(not shown), the spin-spin correlations become increasingly worse with increasing disorder. 
Only for small disorder and interaction strengths do we find reasonable agreement for this quantity.

\begin{table}
\begin{tabular}{ccccccccccccccccc}
(a)      \\[4pt]
&         B      &-0.090 & A & -0.019 & A & -0.129 & B &  -0.025\\[2pt]
&      -0.166  &&     -0.050  &&    -0.133  &&   -0.018\\[2pt]
&         A     & -0.030 & A & -0.094 & B & -0.027 & B &  -0.182\\[2pt]
&      -0.019  &&   -0.096    && -0.077     &&   -0.116\\[2pt]
&         A     &-0.131  &B & -0.075  &A  &-0.036 & A  & -0.033\\[2pt]
&      -0.113  &&   -0.072    &&     -0.143 &&    -0.107\\[2pt]
&         B      &-0.053 & A & -0.156 & B & -0.035 & B & -0.045\\[2pt]
&      -0.025  &&   -0.031    && -0.043     && -0.059\\[10pt]
(b) \\[4pt]
&         B & -0.008 & A &  0.045 & A & -0.048 & B &  0.038\\[2pt]
&     -0.114  &&    0.037  &&   -0.102   &&   0.032\\[2pt]
&         A & -0.078 & A & -0.105 & B & -0.060 & B & -0.125\\[2pt]
&     -0.022  &&   -0.070  &&   -0.072   &&  -0.078\\[2pt]
&         A & -0.053 & B & -0.099 & A & -0.049 & A &  0.031\\[2pt]
&     -0.094  &&   -0.096  &&   -0.115   &&  -0.117\\[2pt]
&         B & -0.023 & A & -0.112 & B & -0.072 & B &  0.030\\[2pt]
&     -0.028  &&    0.009  &&   -0.068   &&  -0.023\\[2pt]
\\
\end{tabular}
\caption{\label{table:SdotSVeq7Ueq7halffilling}
In (a) the exact values of $\langle \Psi | {\hat {\bf S}}_i\cdot {\hat {\bf S}}_j  | \Psi \rangle$
are listed for the $W/t=7$ and $U/t=8$ state at 1/2 filling, wherein the A and B
label the sites of the particular complexion of disorder that we have studied (see the charge
densities shown in the previous table);
(b) shows analogous data, but now calculated within the HF approximation.}
\end{table} 

\section{Discussion}

We have studied the Anderson-Hubbard model, one of the simplest model Hamiltonians that can describe
correlated electrons moving on a disordered lattice. For short one-dimensional chains (up to
a length of 10 sites) and for a 4$\times$4 square lattice, both with periodic boundary conditions,
we have found the exact ground states using the
Lanczos algorithm, as well as the HF states that minimize the variational estimate of the ground-state
energies using product-state trial wave functions. We have allowed for the HF states to have differing 
spin degrees 
of freedom, specifically paramagnetic (non-magnetic), antiferromagnetic with collinear spins, as well
as fully unrestricted HF solutions in which there can be a local moment at every site that points in
any direction that is dictated by the variational principle. There are two possibilities for the
latter case: planar non-collinear spins and non-coplanar non-collinear spins. Using these wave functions
we have compared several static quantities: energies, local charge densities, local spin moments, and 
near-neighbour spin-spin correlations. We have also found the magnitude of the overlap between the exact
and HF solutions.

The HF wave functions seem to produce reasonably good estimates of the ground-state energies, and the
success of this approximation is best at large disorder (or, of course, very small electron-electron 
interactions). Similarly, good overlaps between the exact and HF wave functions are found. 
Perhaps the most impressive success of the HF approximation is
in its ability to reproduce the charge densities that are found in the inhomogeneous ground states, suggesting
that it is doing very well at accounting for the screening of the disorder potential created by the
Hubbard energy $U$ -- we will discuss in a more thorough fashion the lessons learned concerning
screening, from both the Lanczos and HF results for the ground states found of the 4$\times$4 square
lattice, in a future manuscript. Even more impressive agreement of the local charge densities
is found when one uses a partially-projected Gutzwiller wave function, and we will report on the
success of this much more involved and difficult approximation in the following paper \cite{farhoodfar08}.

In contrast to the success of HF in reproducing the charge densities, it is much less successful in its bid 
to represent the magnetic correlations present in the ground-state wave functions. This is not that surprising,
since the HF states are product wave functions and therefore represent all spin-spin correlations as
the products of classical spins, albeit with magnetic moments that can have a magnitude from zero to 1/2.
For example, close to half filling and in the large $U$ limit one should recover Heisenberg-like
effective Hamiltonians, such as the $t-J$ model, and there quantum fluctuations of the moments play
an important role. The HF wave functions, being product states, cannot possibly include such
quantum fluctuations.

Our results did make clear that sometimes allowing for unrestricted HF solutions with non-collinear
spins was important in both lowering the HF estimates of the ground-state energies, and in improving
the ability of HF to reproduce static correlation functions. The appearance of non-collinear spins
may be important in understanding various results in the literature. For example,
the novel metallic phase in two dimensions that was proposed by
Heidarian and Trivedi \cite{heidarian04}. These authors found the metallic phase in approximately the
same region of the disorder/interaction phase diagram where we found non-collinear local moments in
the HF states ($W/t=7$ and 8 and $U/t=8$ at half filling) for the square lattice. Indeed, we also
found indirect evidence for such physics in the exact ground states -- only for $W/t=7$ and $U/t=8$ at half
filling did we find non-singlet ground states -- and for the HF state for these parameters we found a ground
state having a net magnetic moment of one.  For a correlated 
system close to half filling local magnetic moments are formed which have strong antiferromagnetic 
near-neighbour correlations, and as outlined in the detailed work of Shraiman and Siggia the manner in which 
vacancies (say an empty site created by having a doubly occupied site that is close by) become mobile in
such an antiferromagnetic background is by a dipolar backflow of the magnetization current producing
\textit{spin twists} \cite{shraiman88,shraiman89}. If the mobility of the carriers is enhanced by such 
physics it could be related to the predictions of Ref. \cite{heidarian04}.

Work on the \textit{dynamic} properties of interacting and disordered electronic systems may also be affected
by the properties of the local moments that are introduced in the HF approximation. Work by one of us and 
collaborators \cite{fazileh06} used a HF approach with non-paramagnetic but collinear spins, and found 
evidence for a pseudogap-type result that may explain the metal-to-insulator transition found in 
LiTi$_{2-y}$Al$_y$O$_4$ (the Ti sites occupy the sites of a corner-sharing tetrahedral lattice). 
One does not find such a suppression of the density of states at the fermi level without allowing 
for non-paramagnetic states \cite{fazileh08}, and the possible role of the non-collinearity
of the spins is presently being examined. Parallel results were found earlier by Tusch and Logan \cite{tusch93}
for a similar study of the metal-to-insulator (including other possible orderings) for a three-dimensional
simple cubic lattice, but again using a HF product state that restricted the spins to be collinear.
However, the evidence for a suppression of the density of states at the fermi level away from 1/2
filling is much less clear. Based on the results presented here,
whether or not a fully unrestricted spin-invariant HF product state affects such results is presently
being explored. (Indeed, an assessment of whether or not HF is successful in reproducing the dynamical properties
of the exact solutions of the Anderson-Hubbard model may also be possible, utilizing the kind of
approach recently taken elsewhere \cite{chiesa08}.)

\ack
We thank Bill Atkinson and Nandini Trivedi for helpful discussions.
This work was supported in part by the NSERC of Canada and the Hong Kong RGC.

\newpage

\section*{References}

\providecommand{\newblock}{}

\end{document}